\documentclass[3p, preprint]{elsarticle}

\journal{Astroparticle Physics}

\usepackage[english]{babel}
\usepackage{units}
\usepackage{graphicx}
\usepackage{amsmath}
\usepackage{amssymb}
\usepackage{caption}
\usepackage{subcaption}

\usepackage{xcolor}
\usepackage{xspace}

\usepackage{pgf}
\usepackage{caption}
\usepackage{subcaption}

\definecolor{dark-gray}{gray}{0.3}

\newcommand{\Xmax}{$X_{\rm max}$\xspace}

\usepackage{subfig}

\begin{document}

\title{The effect of the atmospheric refractive index on the radio signal of extensive air showers}

\author[ru]{A.~Corstanje\corref{cor}}
\cortext[cor]{Corresponding author}
\ead{A.Corstanje@astro.ru.nl}
\author[ru]{A.~Bonardi}
\author[br,ru]{S.~Buitink}
\author[ru,as,ni,mpifr]{H.~Falcke}
\author[ru,ni]{J.~R.~H\"orandel}
\author[br]{P.~Mitra}
\author[br]{K.~Mulrey}
\author[ru,ni,ir]{A.~Nelles}
\author[ru]{J.~P.~Rachen}
\author[ru]{L.~Rossetto}
\author[ru,pri]{P.~Schellart}
\author[rug,iuibr]{O.~Scholten}
\author[ru]{S.~ter Veen}
\author[ru,vax]{S.~Thoudam}
\author[rug]{G.~Trinh}
\author[br]{T.~Winchen}

\address[ru]{Department of Astrophysics/IMAPP, Radboud University Nijmegen, P.O. Box 9010, 6500 GL Nijmegen, The Netherlands}
\address[br]{Astrophysical Institute, Vrije Universiteit Brussel, Pleinlaan 2, 1050 Brussels, Belgium}
\address[iuibr]{Interuniversity Institute for High-Energy, Vrije Universiteit Brussel, Pleinlaan 2, 1050 Brussels, Belgium}
\address[as]{Netherlands Institute for Radio Astronomy (ASTRON), Postbus 2, 7990 AA Dwingeloo, The Netherlands}
\address[ni]{Nikhef, Science Park Amsterdam, 1098 XG Amsterdam, The Netherlands}
\address[mpifr]{Max-Planck-Institut f\"{u}r Radioastronomie, Auf dem H\"ugel 69, 53121 Bonn, Germany}
\address[ir]{Now at: Department of Physics and Astronomy, University of California Irvine, Irvine, CA 92697, USA}
\address[rug]{University of Groningen, P.O. Box 72, 9700 AB Groningen, The Netherlands}
\address[pri]{Now at: Department of Astrophysical Sciences, Princeton University, Princeton, NJ 08544, USA}
\address[vax]{Now at: Department of Physics and Electrical Engineering, Linn\'euniversitetet, 35195 V\"axj\"o, Sweden}

\begin{abstract}
For the interpretation of measurements of radio emission from extensive air showers, an important systematic uncertainty arises from natural variations of the atmospheric refractive index $n$. 
At a given altitude, the refractivity $N=10^6\, (n-1)$ can have relative variations on the order of $\unit[10]{\%}$ depending on temperature, humidity, and air pressure. Typical corrections to be applied to $N$ are about $\unit[4]{\%}$.
Using CoREAS simulations of radio emission from air showers, we have evaluated the effect of varying $N$ on measurements of the depth of shower maximum \Xmax.
For an observation band of 30 to $\unit[80]{MHz}$, a difference of $\unit[4]{\%}$ in refractivity gives rise to a systematic error in the inferred \Xmax between 3.5 and $\unit[11]{g/cm^2}$, for proton showers with zenith angles ranging from 15 to 50 degrees. 
At higher frequencies, from 120 to $\unit[250]{MHz}$, the offset ranges from 10 to $\unit[22]{g/cm^2}$.
These offsets were found to be proportional to the geometric distance to \Xmax.
We have compared the results to a simple model based on the Cherenkov angle. 
For the $120 - \unit[250]{MHz}$ band, the model is in qualitative agreement with the simulations.
In typical circumstances, we find a slight decrease in \Xmax compared to the default refractivity treatment in CoREAS.
While this is within commonly treated systematic uncertainties, accounting for it explicitly improves the accuracy of \Xmax measurements.

\end{abstract}
\begin{keyword}
Cosmic rays \sep extensive air showers \sep radio emission \sep atmospheric effects
\end{keyword}

\maketitle

\section{Introduction}
In recent years, the techniques for measuring and modelling radio emission from air showers induced by high-energy cosmic rays have developed rapidly \cite{Huege:2016}. 
The radio detection method has achieved high precision in estimating the air shower and primary particle properties \cite{Buitink:2014,Auger:2016,Tunkarex:2016} which allows for very precise measurements of the primary particle type and the energy of the primary cosmic ray \cite{Buitink:2016}.
In particular, the LOFAR radio telescope \cite{van-Haarlem:2013} has been used successfully for this, due to its densely instrumented core region located in the Netherlands. In an inner ring of $\unit[320]{m}$ diameter, we can use 288 low-band antennas, measuring in the $30-\unit[80]{MHz}$ range, for cosmic-ray measurements. Additionally, there are also 288 high-band antennas measuring in the $110$ to $\unit[190]{MHz}$ range.
In an extended core region of about $\unit[6]{km^2}$, nearly 1800 additional low-band antennas have been installed, grouped into stations of 96. Depending on strength and location of the air shower signal, up to four stations outside the inner ring can be used as well.
The signals from air showers have been routinely measured with LOFAR since 2011 \cite{Schellart:2013}.

In an air shower, secondary electrons and positrons are produced, which undergo charge separation as they travel through the Earth's magnetic field. 
This leads to transverse currents producing radio emission. This emission reaches the ground as a short pulse on the order of 10 to $\unit[100]{ns}$ long, with a specific lateral intensity distribution or `footprint' \cite{Nelles:2015} that depends on the depth of shower maximum \Xmax. 
The number of produced secondary particles peaks at \Xmax. This point, expressed as the column density of traversed matter ($\unit{g/cm^2}$), varies with primary particle type and is therefore an important quantity to measure in composition studies.

The measured lateral intensity distributions are compared to microscopic Monte Carlo simulations of air showers, to infer the properties of the primary cosmic ray.
To simulate the radio signal at the antennas we use CoREAS \cite{CoREAS:2013}, a simulation of the radio emission from the individual particles in the cascade simulated with CORSIKA \cite{Corsika:1998}.
Fitting these simulated radio footprints to measured air showers allows us to infer \Xmax to a precision of $\unit[20]{g/cm^2}$. As this precision is comparable to that of fluorescence detection \cite{Auger:2014,Kampert:2012}, it is well suited for composition studies.
The radio detection method is therefore a complementary technique, as it is not limited to dark clear nights, and because its duty cycle is limited only by technical conditions and thunderstorms, it can reach in principle up to $\unit[100]{\%}$.

The detected radio signal depends on the difference in travel times of radio waves and particles. Therefore, it is important to apply an accurate value of the refractive index $n$.
Variations in the refractive index lead to changes in the radio intensity footprint on the ground because the angle of peak emission depends on $n$.
The refractive index, which at sea level is about $n \approx 1.0003$, exhibits natural variations due to weather conditions, at the level of the fifth decimal. It is therefore common to define the refractivity $N=10^6\,(n-1)$, which emphasizes relative variations as these depend on $(n-1)$. 

In this paper, we quantify the influence of the refractive index variations on the depth of shower maximum, to reduce the systematic uncertainty of the \Xmax measurements. In particular, CoREAS currently assumes a constant default value of the refractive index at each altitude in the atmosphere, and we explore how much the systematic error can be reduced by a more accurate treatment of the local atmosphere.
In the next section, we describe a toy model for the radiation from air showers, which qualitatively explains how unaccounted variations in the refractive index give rise to uncertainties in determining \Xmax.
In Sect.~\ref{sect:refractivity} we review the equations used to describe atmospheric parameters and their altitude profiles, and to calculate the refractive index.
In Sect.~\ref{sect:method}, the method of fitting intensity distributions is described, and Sect.~\ref{sect:results} gives the results for the systematic offsets of \Xmax.

\section{Toy model for the effect of the refractive index on radiation from air showers}\label{sect:toymodel}
The depth of shower maximum \Xmax can be inferred from the radio intensity footprint measured on the ground. 
In this section we show the radio footprint changes with the Cherenkov angle, which is a function of the refractive index.
In particular, if the refractive index is higher than expected, the method based on the intensity footprint will underestimate \Xmax.

In an extensive air shower, the magnetic field of the Earth induces an electric current, as the Lorentz force has opposite direction for electrons and positrons in the shower front. This current is transverse to the direction of the shower. 
The number of electrons and positrons depends strongly on the interaction depth of the shower and peaks at \Xmax. The induced current is therefore strongly time dependent and emits electromagnetic radiation. The shower front has a thickness on the order of meters, and the radiation is coherent at wavelengths longer than this thickness, i.e.~at radio wavelengths \cite{Scholten:2012}. 
The transverse current resides in the shower front and thus moves towards the Earth surface with a velocity exceeding the speed of light in air~\cite{Werner:2012}. Therefore, radio waves are emitted because of coherent Cherenkov emission~\cite{deVries:2011}. In addition to the emission from the transverse current there is also a smaller contribution from the net negative charge buildup in the shower front. From the polarization of the radio signal, this contribution was found to be $\unit[11]{\%}$ on average at the LOFAR site \cite{Schellart:2014}.

In the $30 - \unit[80]{MHz}$ band primarily used at LOFAR, the emission along the Cherenkov angle and the non-Cherenkov emission have roughly the same magnitude, while at higher frequencies, the Cherenkov emission dominates (see Fig.~4 in~\cite{deVries:2013}).
This has been confirmed by LOFAR observations above $\unit[110]{MHz}$ where a clear ring-like emission pattern is found~\cite{NellesHBA:2015}. The radius of the ring was on the order of $\unit[100]{m}$ as is expected from the Cherenkov angle and the distance to \Xmax. 
It has also been observed at GHz frequencies by the CROME \cite{CROME:2014} and ANITA \cite{ANITA:2010} experiments.

The angle $\alpha$ where the Cherenkov emission peaks, is given by
\begin{eqnarray}\label{eq:cherenkov}
\cos \alpha & = & \frac{1}{\beta n}, \\
\alpha & \approx & \sqrt{2\,\beta (n-1)}\label{eq:alpha},
\end{eqnarray}
where $\beta = v/c$ is the velocity of the shower front with $\beta=1$ to a good approximation, the refractive index of air $n\approx 1.0003$ at sea level, and it varies with altitude. For convenience we also use the refractivity $N$ throughout the text, as we will consider relative variations in $N$.

The depth of shower maximum \Xmax is reached at an altitude $h_0$, which is given by the altitude-dependent density $\rho(h)$ and the zenith angle $\theta$. The relation is \cite{Auger:2012}
\begin{equation}
X_{\rm max} \equiv X(h_0) = \frac{1}{\cos \theta} \int_{h_0}^\infty \rho(h)\,\mathrm{d}h,
\end{equation}
where $X$ is the column density expressed in $\unit{g/cm^2}$. Therefore, $X$ is also proportional to the pressure,
\begin{equation}\label{eq:Xmax_vs_pressure}
X(h_0) = \frac{10}{g}\,\frac{p(h_0)}{\cos \theta},
\end{equation}
with $g$ the gravitational acceleration and $p$ the pressure in $\unit{Pa}$.

As a toy model for analyzing the effect of varying refractive index, we use the approximation that the size of the radio footprint is proportional to the base of a cone located in the shower plane, with a half-opening angle of $\alpha$ with respect to the the direction of the incoming primary particle. The shower plane is defined as the plane perpendicular to the incoming direction of the primary particle.
Moreover, we assume that all radiation is produced near \Xmax.
As a consequence, the size of the radio footprint on the ground would be proportional to the geometric distance to \Xmax and to the Cherenkov angle at the altitude of \Xmax. 
Variations in refractive index $n$ at altitude $h_0$ would then translate to variations in radio footprint size via the Cherenkov angle.
The radio footprint with its non-circular symmetric structure \cite{Huege:2013,Nelles:2015} falls off smoothly with distance, hence it has no sharply defined ``size". However, the important point for this model is that a given footprint would scale, both with distance to \Xmax and with the Cherenkov angle. 

The model is expected to be more accurate at frequencies above $\unit[100]{MHz}$ where the Cherenkov mechanism dominates. At lower frequencies, the intensity pattern depends less strongly on the Cherenkov angle, and therefore on the refractive index.

As the refractive index is usually discussed in context of optical refraction, it should be noted that the additional effect of variations in $n$ on bending of signal propagation paths (through Snell's law) is negligible for us. A signal path traveling from a medium with $n=1$ to a medium with $n=1.0003$ at a 60 degree incidence angle will be bent by 0.03 degrees. This is already below the resolution of about 0.1 degree attainable with LOFAR \cite{Corstanje:2015}, and natural variations in $n$ are still an order of magnitude smaller.

\begin{figure}
\begin{center}
\includegraphics[width=0.70\textwidth]{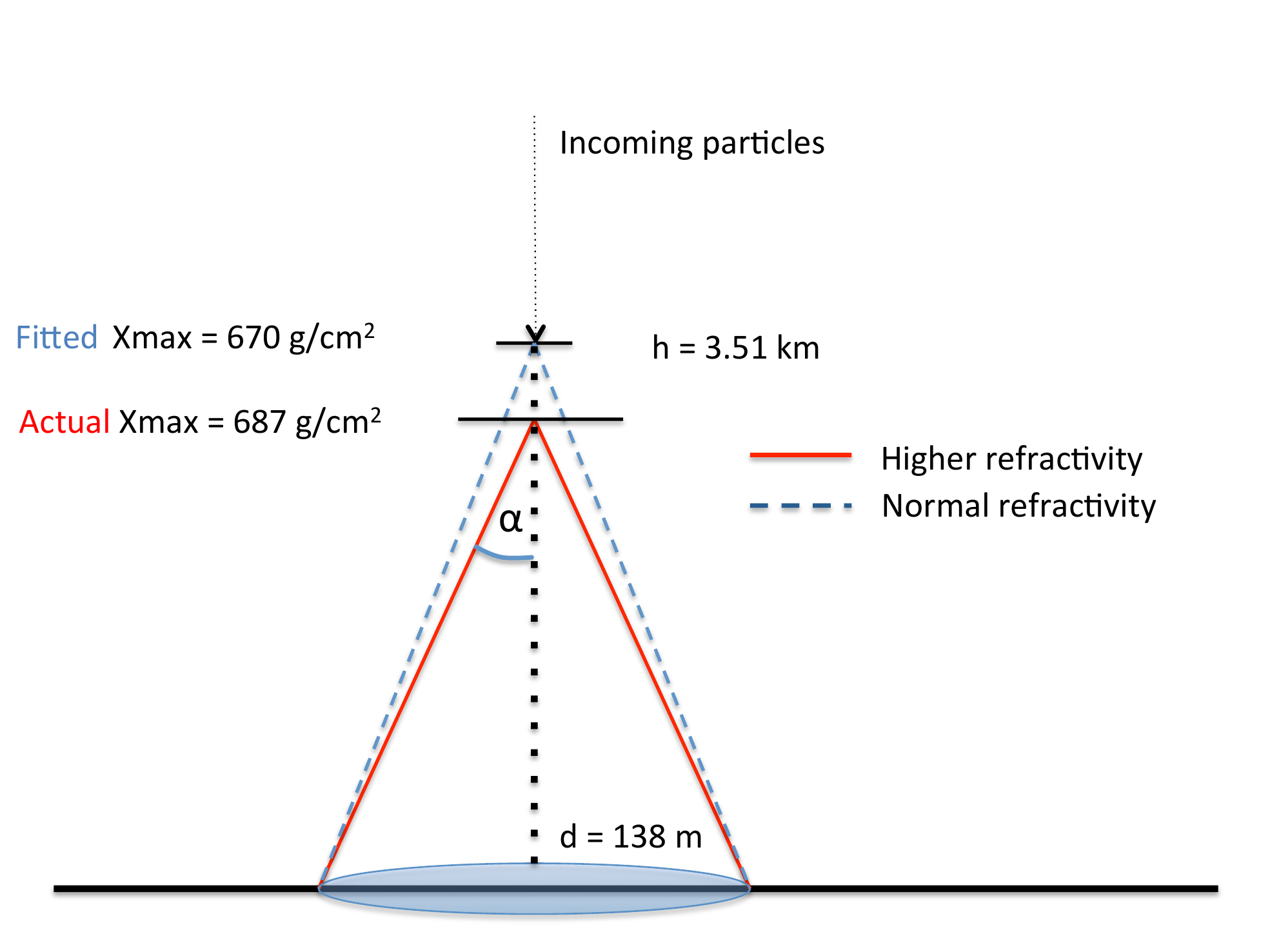}
\caption{Schematic picture (stretched horizontally) of the effect of a higher refractivity on the estimate of \Xmax. This model uses the Cherenkov angle $\alpha$, and the fact that the radio emission is maximal around \Xmax. }
\label{fig:schematic}
\end{center}
\end{figure}

In composition studies, we fit radio intensity profiles simulated with CoREAS to the measured intensity profile on the ground \cite{Buitink:2014}, to estimate \Xmax. In Fig.~\ref{fig:schematic}, the effect of an increase in $n$ is shown schematically, for a proton primary particle of $\unit[10^{17}]{eV}$ from zenith. These have an average \Xmax $\approx \unit[670]{g/cm^2}$ (from CoREAS simulations), corresponding to an altitude of $\unit[3.51]{km}$. 
If the refractive index is higher than expected, fitting the intensity profiles at ground will underestimate \Xmax (blue lines), as for the actual refractive index, the Cherenkov angle is larger (red lines). The lower altitude in the atmosphere corresponds to a higher \Xmax; the modeled difference amounts to $\unit[17]{g/cm^2}$ for a $\unit[10]{\%}$ increase in refractivity $N$. 
This is therefore a systematic uncertainty on \Xmax. 

The offset in \Xmax follows from the condition that the footprint size, which in this approximation is just the diameter of the intersection of the Cherenkov cone with the ground or shower plane, is kept constant. This gives
\begin{equation}\label{eq:heightshift}
\sqrt{N_1 / N_0}\,h_1\;\alpha(h_1) = h_0\;\alpha(h_0),
\end{equation} 
with $h_0$ and $h_1$ the altitude of the \Xmax point at fixed footprint size, for standard refractivity $N_0$ and $N_1=1.1\,N_0$ respectively. 
The square root arises from the small-angle approximation in Eq.~\ref{eq:alpha}; we have also taken $\sin \alpha \approx \alpha$.  It holds both for vertical and for inclined showers and takes into account that $N$, and therefore $\alpha$, varies with altitude as shown in the equations in the next section.

In summary, the steps for calculating the modelled shift in \Xmax given an original \Xmax at altitude $h_0$, are:
\begin{itemize}
\item establish $N_0$ at altitude $h_0$ using the equations in the next section (Eqs.~\ref{eq:temperature} through \ref{eq:refractivity}). 
\item consider a different refractivity profile with altitude, e.g.~$N_1(h) = 1.1\,N_0(h)$
\item solve $h_1$ from Eq.~\ref{eq:heightshift}
\item obtain the column density (shifted \Xmax) at this height, from Eq.~\ref{eq:Xmax_vs_pressure} and Eq.~\ref{eq:pressure} below
\end{itemize}

In the example of Fig.~\ref{fig:schematic}, we would have $h_0=\unit[3.51]{km}$, for which $N_0=192.3$. 
When considering a $\unit[10]{\%}$ higher refractivity, i.e.~$N_1(h) = 1.1\,N_0(h)$ at all altitudes, we obtain $h_1=\unit[3.31]{km}$ when keeping the footprint size constant. At this level, we would have a column density $X=\unit[687]{g/cm^2}$, which is a systematic offset of $\unit[17]{g/cm^2}$.

\section{The atmospheric refractive index}\label{sect:refractivity}
The refractive index $n$ and refractivity $N$ vary with temperature, pressure, and humidity in the atmosphere.
As these depend on altitude, we use a parametrization for pressure and temperature corresponding to the US Standard atmosphere \cite{US:1976}, where we have rewritten the equations to a slightly more compact form. We use  only the bottom layer of this model, from 0 to $h=\unit[11]{km}$, where $h=0$ defines sea level.
It is valid in the troposphere under `average' circumstances, 
\begin{eqnarray}\label{eq:USstandard}
T(h) & = & T_0 - L h, \label{eq:temperature} \\
p(h) & = & p_0 \left(\frac{T(h)}{T_0}\right)^{\frac{g M}{L R}}\label{eq:pressure}.
\end{eqnarray}
Here, $p_0 = \unit[1013.25]{hPa}$ and $T_0=\unit[288.15]{K}$ are standard sea level pressure and temperature. These can be varied according to local circumstances.
The temperature lapse rate $L$ is assumed constant at $L = \unit[6.5]{K / km}$. 
The remaining constants are the gravitational acceleration $g=\unit[9.80665]{m/s^2}$, the ideal gas constant $R=\unit[8.31447]{J/(mol\,K)}$, and the molar mass of dry air $M = \unit[0.0289644]{kg/mol}$.
In the limit $L \rightarrow 0$, Eq.~\ref{eq:pressure} reduces to the familiar exponential barometric formula.

Given the relative humidity $H$, the partial pressure of water vapor is calculated using the Magnus formula for the saturation pressure \cite{Buck:1981}:
\begin{eqnarray}\label{eq:sat_pressure}
p_{\rm sat} & = & a \exp\left(\frac{b\,t}{t + c}\right) \\
p_w & = & H\, p_{\rm sat},
\end{eqnarray}
where $t$ is temperature in $\unit{^\circ C}$ (in contrast to $T$ in K), and $p_{\rm sat}$ is the saturation pressure for water vapor. For the empirically determined constants we take the values from \cite{Buck:1981}, which are $a=\unit[6.1121]{hPa}$, $b=17.502$, $c=\unit[240.97]{K}$ for temperatures above $\unit[0]{^\circ C}$, and $a=\unit[6.1115]{hPa}$, $b=22.452$, $c=\unit[272.55]{K}$ below $\unit[0]{^\circ C}$. The relative uncertainty in $p_{\rm sat}$ is then given as $\unit[0.1]{\%}$ over the range of $-50$ to $\unit[+40]{^\circ C}$ .

The radio refractivity is parametrized according to \cite{Rueger:2002} as
\begin{equation}\label{eq:refractivity} 
N = 77.6890\, \frac{p_d}{T} + 71.2952\, \frac{p_w}{T} + 375463\, \frac{p_w}{T^2},
\end{equation}
where $T$ is temperature in K; $p_d$ and $p_w$ are the partial pressures (hPa) of dry air and of water vapor, respectively. The total air pressure is $p = p_w + p_d$.
The small influence of carbon dioxide is included in the dry-air contribution. The accuracy of this equation is given as $\unit[0.02]{\%}$ for the first term in Eq.~\ref{eq:refractivity}, and $\unit[0.2]{\%}$ for the second and third term combined. 
This evaluates to a relative uncertainty of less than $\unit[0.1]{\%}$ in total, which is sufficient for our purposes.

It should be noted that at the radio frequencies of interest here, the refractivity values are different from those at infrared, visible, and UV wavelengths, such as considered in \cite{Auger:2012} for the fluorescence detection technique at Pierre Auger Observatory. In particular, the presence of water vapor significantly raises the radio refractivity, while it tends to lower the infrared refractivity slightly below that of dry air. The latter is depicted in \cite{Bernlohr:2014} regarding optical Cherenkov telescopes.
Therefore, the precision formulas of Edl\'en (updated in \cite{Birch:1993}) and Ciddor \cite{Ciddor:1996}, defined for visible and near-infrared wavelengths are not applicable for us, and one can also not simply take the infinite-wavelength limit of those. 
For instance, for air at standard pressure, $\unit[20]{^\circ C}$, and $\unit[50]{\%}$ relative humidity, the Ciddor equation gives $N = 268$, while Eq.~\ref{eq:refractivity} gives $N = 319$. It follows that for radio detection, accounting for humidity is more important than for fluorescence detection.

With the above formulas and definitions, we calculate the atmospheric profile of the refractivity versus altitude for different values of temperature, pressure, and humidity. 
In the US Standard atmosphere one uses the geopotential altitude, which takes into account the decrease in gravitational acceleration $g$ with altitude. As the difference between geometric and geopotential altitudes at $h=\unit[5]{km}$ is only 4 m, we do not correct for the difference.

\begin{figure}
\begin{center}
\includegraphics[width=0.70\textwidth]{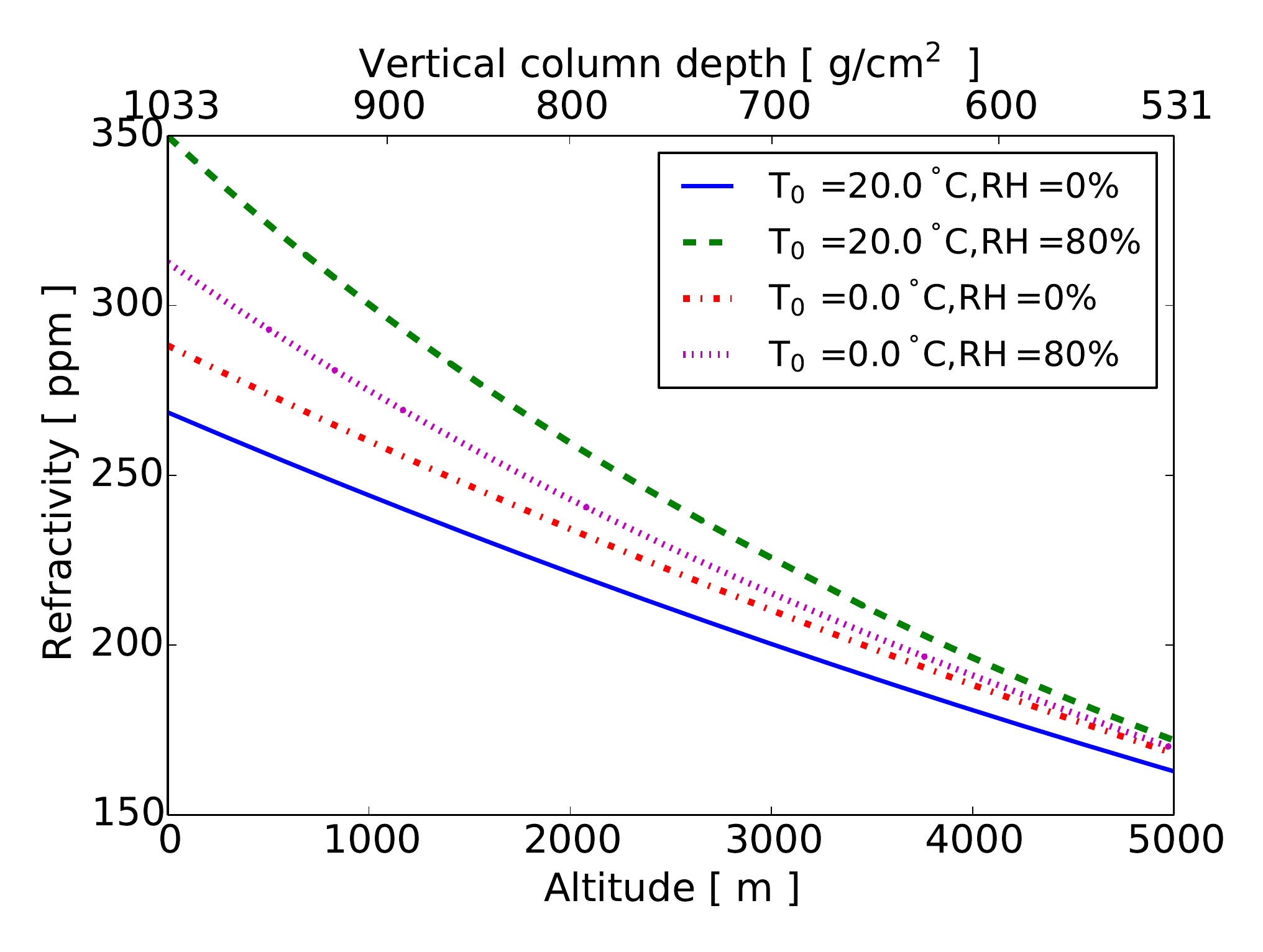}
\caption{Four example altitude profiles of refractivity from Eqs.~\ref{eq:refractivity} and \ref{eq:USstandard}, all assuming standard pressure at sea level. 
For two values of sea-level temperature $T_0$, dry air is compared to a more realistic relative humidity (RH) in the Netherlands, around $\unit[80]{\%}$.}
\label{fig:refractivity_profile}
\end{center}
\end{figure}

\begin{figure}
\begin{center}
\begin{subfigure}{0.70\textwidth}
\includegraphics[width=\textwidth]{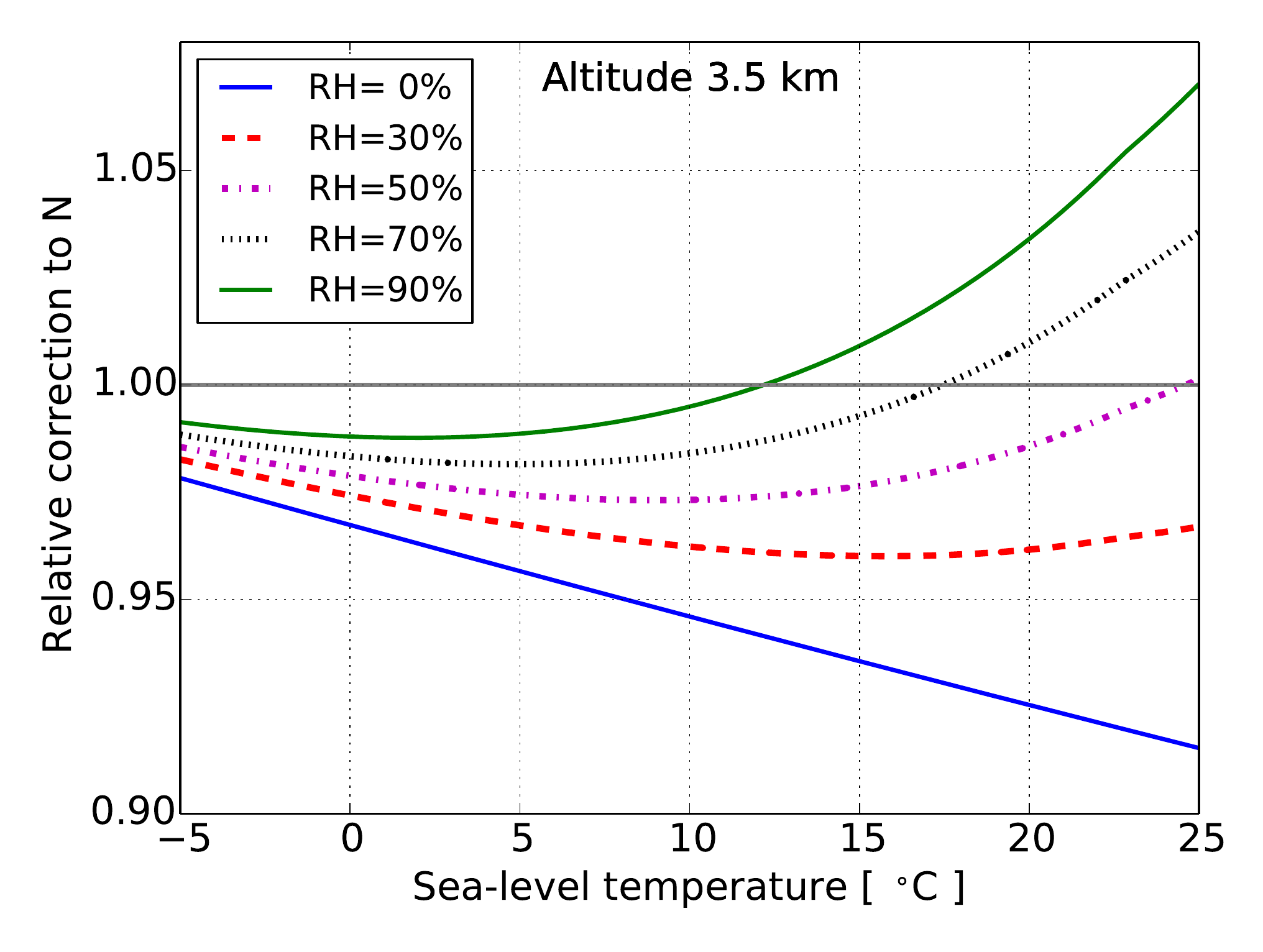}
\caption{}
\label{fig:refractivity_correction_3500}
\end{subfigure}
\begin{subfigure}{0.70\textwidth}
\includegraphics[width=\textwidth]{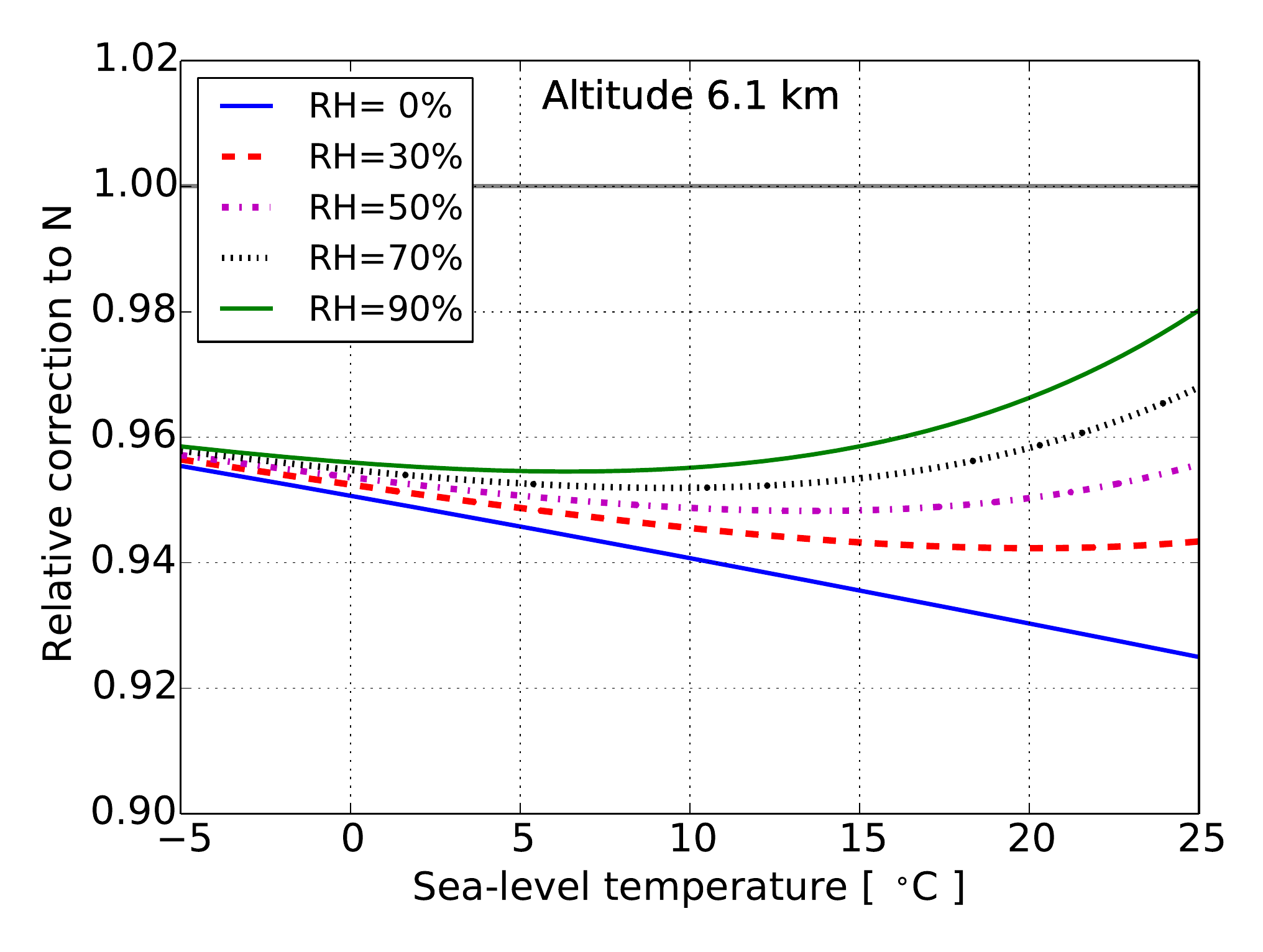}
\caption{}
\label{fig:refractivity_correction_6100}
\end{subfigure}
\caption{\label{fig:refractivity_correction}%
Relative correction to the standard CoREAS refractivity, as a function of ground-level temperature and relative humidity at the given altitude. The altitudes corresponds the average \Xmax of a $\unit[10^{17}]{eV}$ proton shower, (a) vertical, and (b) at a 45 degree zenith angle. In case (b), the correction profiles are always below unity in this temperature range.} 
\end{center}
\end{figure}

An example plot of the altitude dependence of $N$ is shown in Fig.~\ref{fig:refractivity_profile} for two values of temperature and humidity. It is clear that humidity cannot be neglected, especially in the Netherlands where relative humidity near sea level is on average roughly $\unit[80]{\%}$ \cite{klimaatatlas}.
In CoREAS, the refractivity is set to a default value of $N=292$ at sea level, and scaled with density according to the US Standard atmosphere, which considers dry air and $T=\unit[15]{^\circ C}$ at sea level.
In Fig.~\ref{fig:refractivity_correction_3500}, we show the relative correction factor to $N$ with respect to its default value, at an altitude of $\unit[3.5]{km}$. This corresponds to \Xmax = $\unit[670]{g/cm^2}$, which is the average value found for protons of $\unit[10^{17}]{eV}$ energy.
The correction is plotted against the ground temperature for several values of the relative humidity at the given altitude. 
At a sea-level temperature of $\unit[10]{^\circ C}$ and $\unit[70]{\%}$ relative humidity, the true refractivity is about $\unit[2]{\%}$ lower than the default value. At higher sea-level temperatures, humidity plays a larger role, and the curves show a larger spread.

For air showers coming in at a 45 degree zenith angle, the altitude of \Xmax is about $\unit[6.1]{km}$.
At this altitude, the actual $N$ is around $\unit[5]{\%}$ lower than the default value.

For the air pressure, we have taken the standard value of $p=\unit[1013.25]{hPa}$.
Natural variations, which are on the order of $\pm \unit[2]{\%}$, have the effect of lowering or raising the altitude of a given column density \Xmax. 

The relative humidity is expected to vary with altitude, dependent on conditions like cloud cover. 
The humidity in a region around the \Xmax altitude will be the most important. Therefore, to accurately estimate $N$, it is necessary to use atmospheric profile data such as available e.g.~through the Global Data Assimilation System (GDAS) \cite{GDAS}. Uncertainties in these data translate into the  uncertainty in $N$. Ref.~\cite{Auger:2012} gives a comparison between GDAS data and weather balloon measurements in Argentina.
Over the years 2009 and 2010, and in the altitude range of 3.5 to $\unit[6]{km}$, this difference is $\pm \unit[0.5]{^\circ C}$ for temperature, $\unit[0.5]{hPa}$ for pressure, and $\unit[0.05]{hPa}$ for water vapor pressure. The latter corresponds to about $\unit[2]{\%}$ relative humidity at $\unit[3.5]{km}$ altitude, and about $\unit[7]{\%}$ at $\unit[6]{km}$.

This shows that the GDAS data accurately represent the local circumstances. Although the uncertainty numbers may be different for other experiment sites, they are expected to be on the same order.
The resulting relative uncertainty in $N$ is around $\unit[0.5]{\%}$ and is dominated by the humidity uncertainty. This is sufficient for our purposes.

\section{Method}\label{sect:method}
To investigate the effect of changing the refractive index on \Xmax measurements, we considered proton showers with a primary energy of $\unit[10^{17}]{eV}$ for five different zenith angles. For each case, we have generated an ensemble of 50 simulated showers with $N$ at its default value of $N=292$ at sea level.
We have also generated another set of 50 showers where the refractivity $N$ is higher by $\unit{10}{\%}$ at all altitudes. 
This is done using the same random number seeds, ensuring the evolution of the particles is exactly the same. Only the radio emission is recalculated, taking into account the higher refractivity.
This method is similar to the one used in \cite{Buitink:2014} that was used for the composition analysis at LOFAR \cite{Buitink:2016}. The difference is that here we compare two simulated ensembles instead of comparing simulations to measured data.

For all showers, we evaluate the lateral distribution of signal intensity in the shower plane $f(x, y)$, for a star-shaped pattern of antennas. In this pattern, 160 antennas are laid out along 4 lines, at a distance $\unit[25]{m}$ apart. The lines intersect at the origin and make angles of 45 degrees to each other, forming an octagonal star pattern.
For the signal intensity we use a bandpass filter to limit the frequency range to $30-\unit[80]{MHz}$, relevant for the LOFAR cosmic-ray project. 
For comparison, we also consider the $\unit[120-250]{MHz}$ band, where Cherenkov effects are more important.

We then take one of the showers with higher refractivity as a `test shower', and fit the lateral distribution of each of the 49 other showers at the default refractivity to it.
When fitting the shower with index $k$ to the test shower, this yields a mean-square error as a fit quality measure:
\begin{equation}\label{eq:fitquality}
\mathrm{MSE}(k) = \frac{1}{N_{\rm ant}} \sum_{\rm antennas} \left(A f_{\rm test}(x, y) - f_k(x, y) \right)^2,
\end{equation}
where ($x$, $y$) are the antenna positions in the shower plane, and $A$ is a scale factor that is taken as a free parameter. This is proportional to a reduced $\chi^2$ for the case where all antennas have the same uncertainty on the intensity. Additionally, as we simulate antennas in a star-shaped pattern, we apply weight factors such that each antenna represents the same amount of area in the footprint.
The optimal value of $A$ that minimizes the MSE, is given by
\begin{equation}
A = \frac{\sum f_{\rm test}(x, y)\,f_k(x, y)}{\sum f_{\rm test}(x, y)^2}.
\end{equation}

\begin{figure}
\begin{center}
\includegraphics[width=0.70\textwidth]{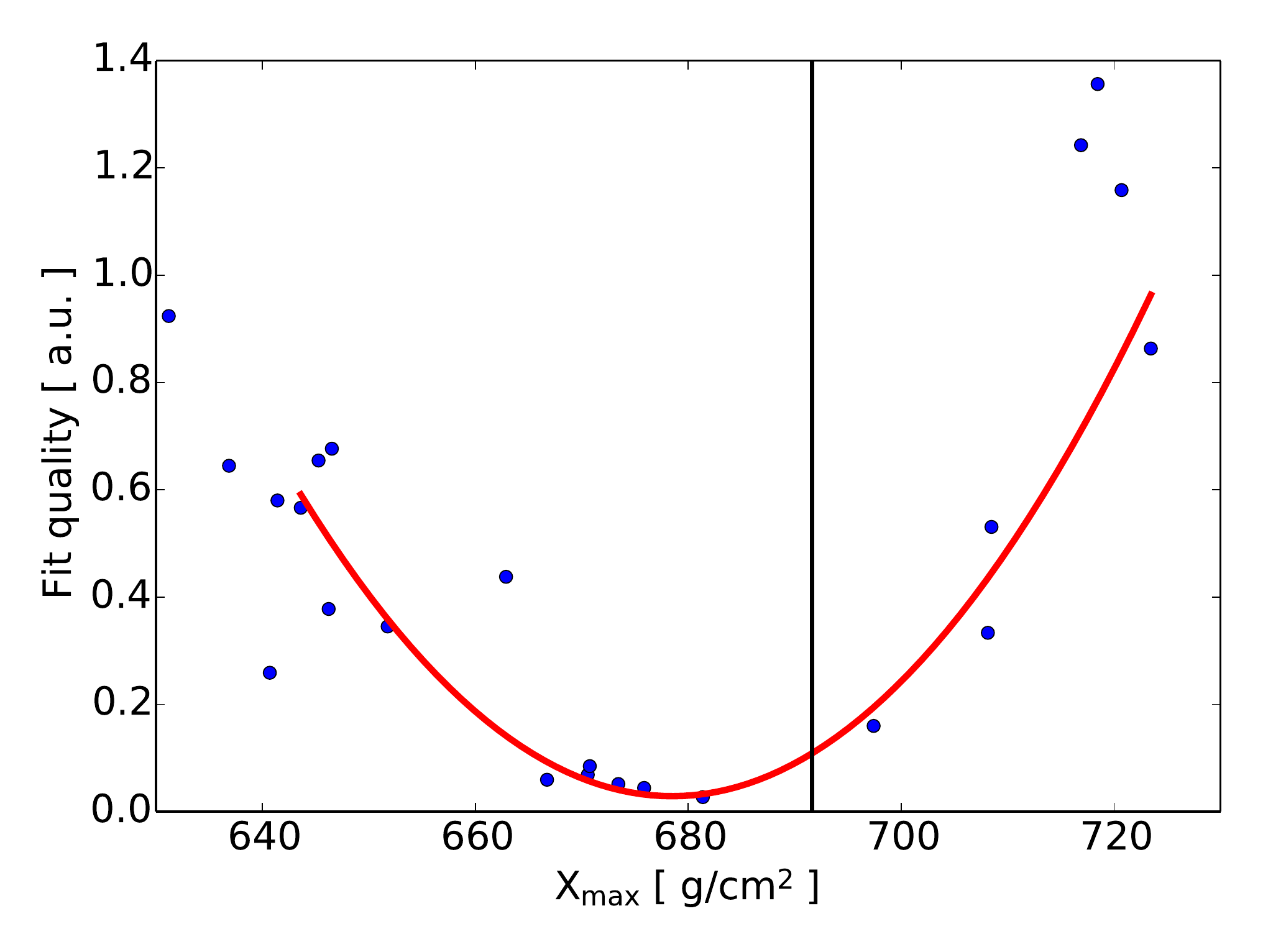}
\caption{An example result of the fitting procedure of lateral distributions. This plot corresponds to one shower simulated at higher refractivity, fitted by an ensemble of showers at standard refractivity to evaluate the shift in \Xmax. The simulated \Xmax is indicated by the black vertical line. Every dot corresponds to the fit to this shower of one shower from the ensemble simulated at normal refractivity. 
The fit qualities as defined in Eq.~\ref{eq:fitquality}, when fitted by a parabola, show a minimum at $\unit[679]{g/cm^2}$, which can be compared to the originally simulated \Xmax value.}
\label{fig:fit_parabola}
\end{center}
\end{figure}

For every shower, we plot the fit quality (MSE) versus \Xmax. It is expected to have a minimum, and to lowest order, to have a quadratic dependence around the minimum. Therefore, we fit the points with a parabola.
The location of the minimum of the parabola is used as an estimator for \Xmax. For one shower this is shown in Fig.~\ref{fig:fit_parabola}, for a limited range around the minimum. 
The scatter around the fitted parabola arises from natural shower-to-shower fluctuations.

We found it useful to weight the fit-quality data points like shown in Fig.~\ref{fig:fit_parabola} by their inverse square in the parabolic fit, thus putting more emphasis on well-fitting profiles. This lowers the uncertainty in estimating the minimum \Xmax by up to $\unit[20]{\%}$, without introducing a bias. 
Proceeding this way, the inferred \Xmax of the test shower with higher refractive index has a systematic offset with respect to its true \Xmax value.
By taking in turn each of the 50 showers in the ensemble as test shower, this offset is evaluated along with its statistical uncertainty. 
This quantifies the effect of variations in the refractive index in e.g.~a composition analysis where simulations are fitted to data.

We have limited the range of the fit to include those values of \Xmax within $\pm \unit[40]{g/cm^2}$ of the expected value.
As this range is determined by the true \Xmax and the offset, we shift the fit range in a second iteration. The offsets found in both iterations are consistent.
We have discarded 5 showers at the low, and 5 at the high end of the true \Xmax range, as these give less accurate parabola fits due to lack of data points at one side of the range.

The parabolic fit procedure has also been tested using only showers with the same refractive index. On average over 50 showers, the true \Xmax is reproduced within standard errors. For the ensemble of 50 minus 10 showers, the standard error ranges from 1 to $\unit[2]{g/cm^2}$ for zenith angles from 15 to 50 degrees, respectively.

\section{Results}\label{sect:results}
\begin{figure}
\begin{center}
\begin{subfigure}{0.70\textwidth}
\includegraphics[width=\textwidth]{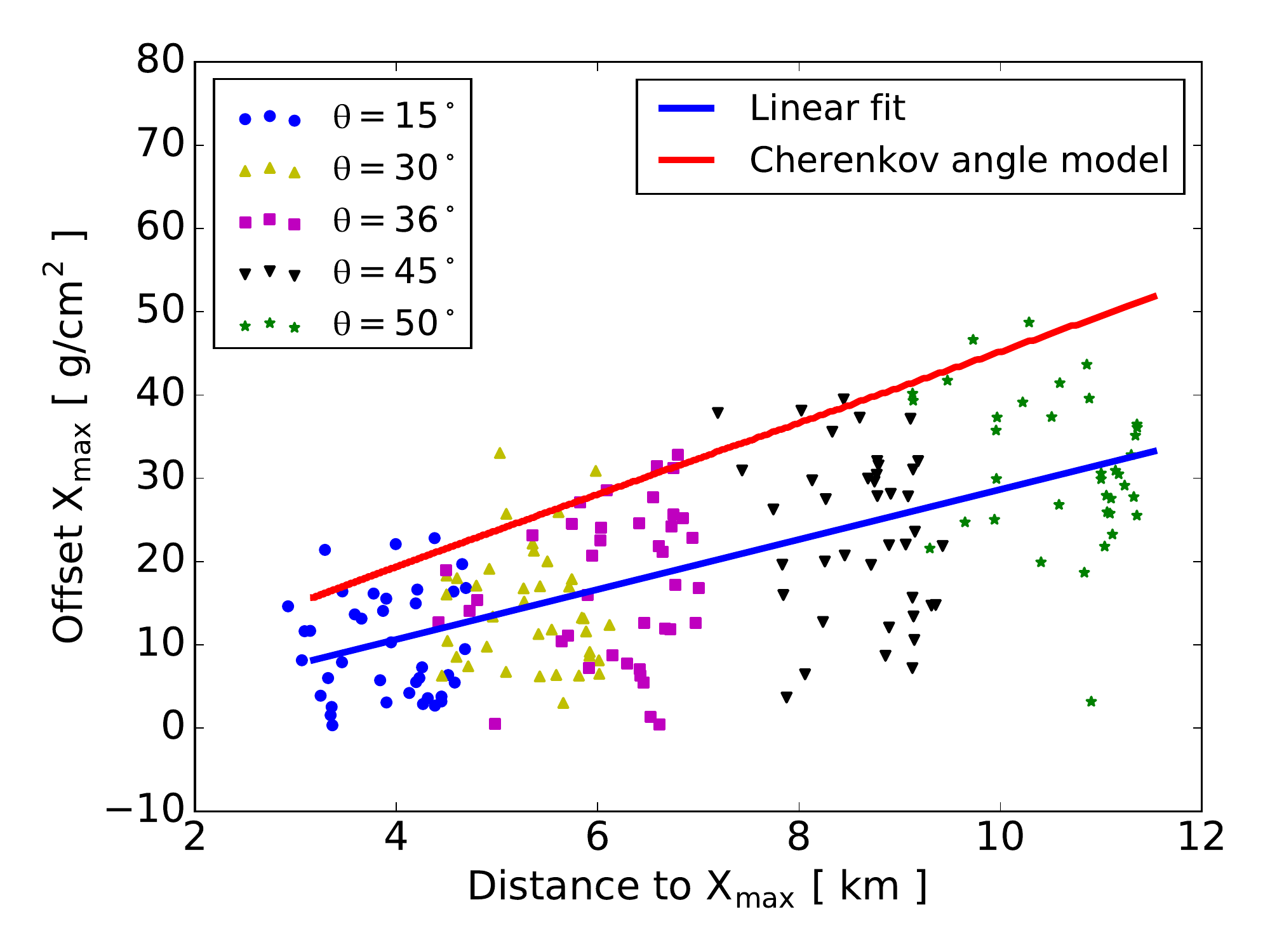}  
\caption{}
\label{fig:xmax_offset_LBA}
\end{subfigure}
\begin{subfigure}{0.70\textwidth}
\includegraphics[width=\textwidth]{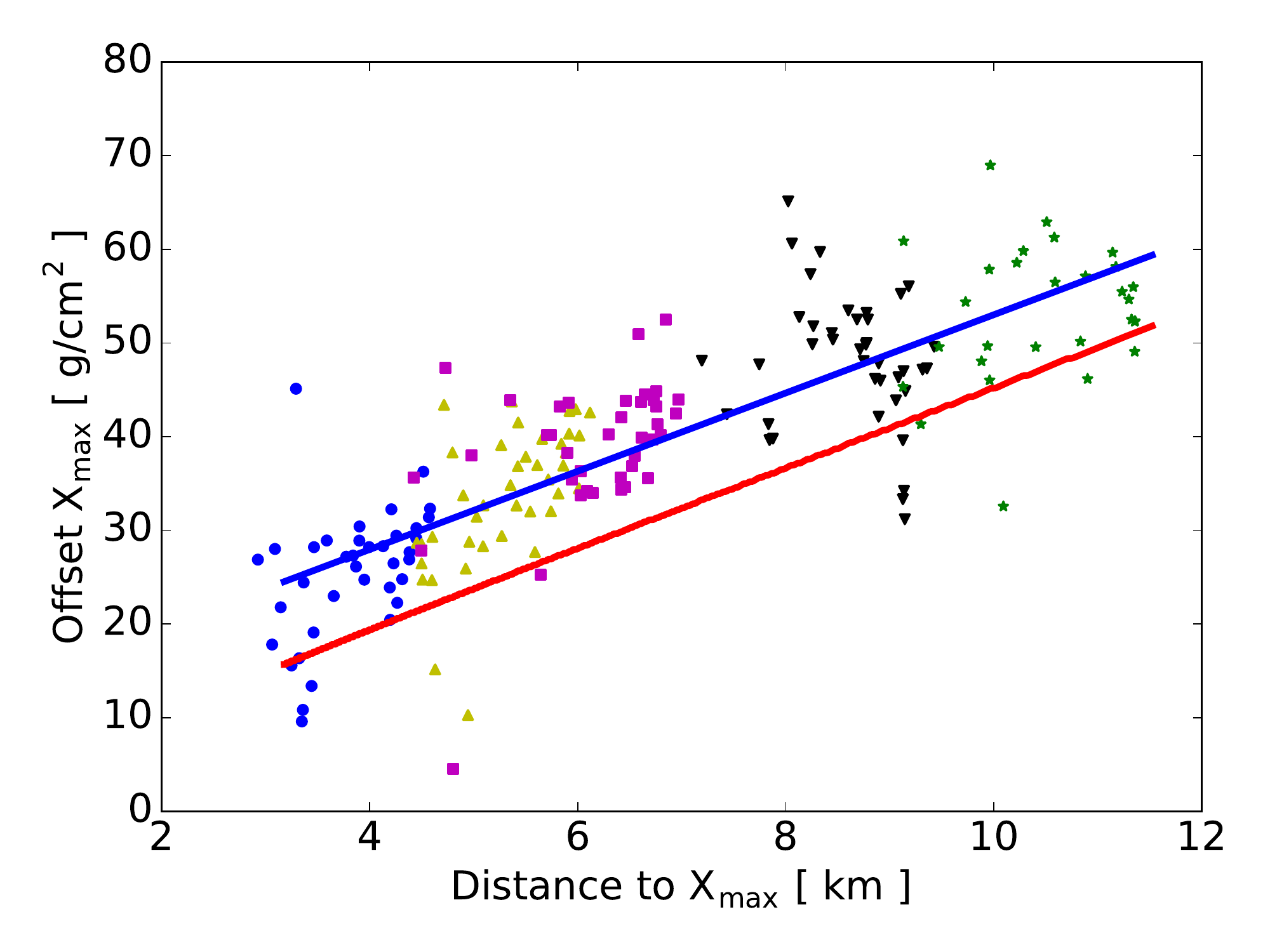} 
\caption{}
\label{fig:xmax_offset_HBA}
\end{subfigure}
\caption{The \Xmax offset for zenith angles of 15, 30, 36, 45, and 50 degrees respectively, where the points represent individual simulated showers. The blue line is a linear fit to the data points. In (a), for the 30 to $\unit[80]{MHz}$ range, the simplistic model (red line) gives nearly twice the offsets found in simulations. For the 120 to $\unit[250]{MHz}$ range shown in (b), the simulated offsets are about $\unit[10]{g/cm^2}$ below the simulations.  }
\end{center}
\end{figure}

We have evaluated the bias in \Xmax for a $\unit[10]{\%}$ increase in refractivity, for five different zenith angles between 15 and 50 degrees.
An increase in $N$, when not taken into account, leads to an overestimate in the geometric distance to \Xmax, and hence to an underestimate in \Xmax.
Making the simplistic assumption that all radiation is produced at \Xmax, and that the size of the radio footprint scales with the Cherenkov angle (Eq.~\ref{eq:cherenkov}), we apply the procedure given in Sect.~\ref{sect:toymodel} to find the shift in \Xmax.

For the $30-\unit[80]{MHz}$ band, the results from the simulation fits are plotted together with the model predictions in Fig.~\ref{fig:xmax_offset_LBA}. A linear fit to the results for individual showers is also shown (blue line).
The modeled offset in \Xmax is nearly linear in the distance to \Xmax, and is a bit less than twice the offsets from simulations. The modeled offset reproduces the general dependence on geometric distance, but the simulated offsets are about a factor 2 smaller. As explained in Sect.~\ref{sect:toymodel}, this is understood, as for low frequencies the non-Cherenkov emission is about equally important as the Cherenkov emission, leading to a weaker dependency on $N$.
The offset ranges from about $\unit[9]{g/cm^2}$ for a zenith angle of 15 degrees, to $\unit[28]{g/cm^2}$ at $\theta=50$ degrees.
A linear fit suffices to describe the data; a higher-order polynomial curve does not significantly reduce the residuals. The standard deviation of the residuals is $\unit[8.9]{g/cm^2}$, which is also the intrinsic uncertainty in this fit method, due to natural shower-to-shower fluctuations.
The relation for the shift $\Delta X$, defined as the underestimation of \Xmax per $\unit[10]{\%}$ increase in refractivity, is
\begin{equation}\label{eq:shift_lba}
\Delta X_{\unit[10]{\%}} = 3.00\, \left(\frac{R}{\unit[1]{km}}\right) - 1.37\; \unit{g/cm^2},
\end{equation}
with $R$ the geometric distance to \Xmax. It can be scaled with the relative change in refractivity.

For the $120-\unit[250]{MHz}$ band, the \Xmax offset is shown in Fig.~\ref{fig:xmax_offset_HBA}.
Here, the simulated offsets are well reproduced by the model, up to an almost constant additional shift.
The fitted linear function for the offset is
\begin{equation}\label{eq:shift_hba}
\Delta X_{\unit[10]{\%}} = 4.18\, \left(\frac{R}{\unit[1]{km}}\right) + 11.24\; \unit{g/cm^2}.
\end{equation}
The standard deviation of the fit residuals is $\unit[7.2]{g/cm^2}$ which is again not significantly reduced when using a quadratic function.

The relative correction to the refractivity from Fig.~\ref{fig:refractivity_correction} is almost always below unity for vertical showers, and always below unity for inclined showers in the temperature range relevant at LOFAR.
Therefore, this gives a systematic offset, not only in individual \Xmax estimates, but also in the average \Xmax inferred from many air showers.

Over the range $-5$ to $\unit[20]{^\circ C}$, the actual $N$ is about $\unit[1.5]{\%}$ below the default simulated value for vertical showers. 
For inclined showers, the actual $N$ is about $\unit[4.5]{\%}$ below the default.
Hence, the average \Xmax is overestimated by about $\unit[1.5]{g/cm^2}$ for near-vertical showers, and by about $\unit[11]{g/cm^2}$ at $\unit[45]{degrees}$ inclination. 

The bias is larger for more inclined air showers; it has been noted in the LOFAR composition study \cite{Buitink:2016} that more inclined air showers (above 32 degrees zenith angle) had on average a higher inferred \Xmax than the more vertical ones. The difference in average \Xmax amounts to $\unit[16]{g/cm^2}$ between the two cases.
From the offsets in Eq.~\ref{eq:shift_lba}, it follows that at most $\unit[9]{g/cm^2}$ of this can be explained by a bias caused by using an incorrect value of the refractive index.
 
The fact that $N$ is slightly overestimated in simulations is also relevant for other radio detection experiments such as AERA \cite{Auger:2016} and Tunka-Rex \cite{Tunkarex:2016}. 
These experiments, located in Argentina and Siberia respectively, have a temperature range different from the LOFAR site. Nevertheless, for the case of inclined showers, the relative correction as from Fig.~\ref{fig:refractivity_correction_6100} is below unity down to well below $\unit[-40]{^\circ C}$. 

The numerical constants in Eqs.~\ref{eq:shift_lba} and \ref{eq:shift_hba} follow from the simulations, which depend on location parameters such as height above sea level and the geomagnetic field vector. Therefore, for other experiments the bias on the average \Xmax is expected to be on the same order, but with slight differences due to variations in location-specific parameters, leading to different constants in Eqs.~\ref{eq:shift_lba} and \ref{eq:shift_hba}.

\section{Summary}
The technique of measuring radio signals from air showers to infer the mass composition of cosmic rays, relies on accurate measurements of the depth of shower maximum \Xmax. One of the systematic uncertainties on \Xmax is given by the refractive index of air in the atmosphere, which exhibits natural variations. 

We have evaluated the effect of variations in the refractive index of air on determining \Xmax. 
Using a procedure similar to that used in the composition study at LOFAR \cite{Buitink:2016}, we have taken simulated proton showers at a $\unit[10]{\%}$ higher refractivity $N=10^6\,(n-1)$. We have fitted them with a Monte Carlo ensemble of 49 showers simulated at a default value of $N$, excluding the one corresponding to the higher-$N$ shower being fitted.
The minimum in the least-squares fit quality yields, on average over many showers, the systematic offset in \Xmax.

These offsets were found to be proportional to the geometric distance to \Xmax, for zenith angles ranging from 15 to 50 degrees.
The effect is roughly twice as strong for the $120-\unit[250]{MHz}$ band as for the $30-\unit[80]{MHz}$ band.
Given variations in $N$ on the order of $\unit[4]{\%}$, from the atmospheric effects described in Sect.~\ref{sect:refractivity}, 
the offsets would range from about $3.5$ to $\unit[11]{g/cm^2}$ for $30-\unit[80]{MHz}$, and from $10$ to $\unit[22]{g/cm^2}$ for $120-\unit[250]{MHz}$.

A simplistic model in which the radio intensity pattern is assumed to be proportional in size to the Cherenkov cone starting from the \Xmax point, qualitatively describes the effect; the fitted offsets were found to be just above half the modeled offsets in the $30-\unit[80]{MHz}$ band, and about 20 \% above the modeled offsets in the $120-\unit[250]{MHz}$ band.

Calculated profiles of refractivity versus altitude show that one cannot use a single default value of $N$ in simulations.
This leads to a bias in the average \Xmax, depending on the choice of a constant $N$ either for near-vertical or for inclined showers.
The accuracy can be improved for individual showers by using Eqs.~\ref{eq:shift_lba} and \ref{eq:shift_hba}.

As a further improvement, it would be required to include detailed atmospheric data, including the particular refractivity profile at the time of the air shower, into a next version of Corsika / CoREAS. The GDAS database \cite{GDAS} is useful for this. 
This allows to fully account for the refractive index variations, and to re-evaluate the LOFAR measurements at the best level of detail. This is the subject of a future publication.

\section*{Acknowledgements}
We acknowledge financial support from the Netherlands Organization for Scientific Research (NWO), VENI grant 639-041-130, the Netherlands Research School for Astronomy (NOVA), the Samenwerkingsverband Noord-Nederland (SNN) and the Foundation for Fundamental Research on Matter (FOM). We acknowledge funding from the European Research Council under the European Union's Seventh Framework Program (FP/2007-2013) / ERC (grant agreement n. 227610) and under the European Union's Horizon 2020 research and innovation programme (grant agreement n. 640130).

\bibliographystyle{elsarticle-num}
\bibliography{refractivity}

\end{document}